\documentclass[11pt]{article}
\usepackage[margin=2cm, top=2cm, bottom=2.5cm,a4paper]{geometry}
\usepackage{amsmath}
\usepackage{amssymb}
\usepackage{float}
\usepackage{bbding}
\usepackage{authblk}
\usepackage{bbm}
\usepackage{comment} 
\usepackage{fancyhdr} 
\usepackage{fancyvrb} 
\usepackage{empheq}
\usepackage{graphicx} 
\usepackage{listings} 
\usepackage{mathtools}
\usepackage[version=3]{mhchem}
\usepackage{caption, threeparttable}
\usepackage{paralist} 
\usepackage{url} 
\usepackage[capitalize,nameinlink,noabbrev]{cleveref}

\usepackage{newtxtext}

\def\be{\begin{equation}}
\def\ee{\end{equation}}
\def\bes{\begin{equation*}}
	\def\ees{\end{equation*}}
\def\bea{\begin{equation} \begin{aligned}}
\def\eea{\end{aligned} \end{equation}}
\def\beas{\begin{equation*} \begin{aligned}}
		\def\eeas{\end{aligned} \end{equation*}}

\def\bi{\begin{itemize}}
	\def\ei{\end{itemize}}

\begin{document}
\title{Nonequilibrium ensemble averages using nonlinear response relations}

\author[1]{Manuel Santos Guti\'errez \thanks{Email: \texttt{m.santosgutierrez@leicester.ac.uk}}}
\author[1]{Valerio Lucarini\thanks{Also at: School of Sciences, Great Bay University, Dongguan, P.R. China.}} 
\author[1]{John Moroney} 
\author[1]{Niccolò Zagli} 
\affil[1]{School of Computing and Mathematical Sciences, University of Leicester, Leicester, LE1 7LE, UK.}

\maketitle

\begin{abstract}
The transient time correlation function (TTCF) method is widely used in molecular fluids to compute non-equilibrium transport quantities, providing improved signal-to-noise ratios in ensemble averages without requiring prohibitively large sample sizes. In spite of its success in molecular and turbulent fluid systems, the method has not been systematically explored for more general non-equilibrium dynamical systems, including geophysical applications where the invariant measure is typically unknown. In this work, we present an analytical and numerical investigation of the TTCF method for computing nonlinear response functions in systems far from equilibrium. We discuss its relation to the spectral theory of stochastic systems, highlighting regimes where linear theory is insufficient and the advantages of TTCF. The aim of this work is to provide a framework for studying transient and steady-state responses using the TTCF approach in a broad class of nonequilibrium systems. 
\end{abstract}

\section{Introduction}

Computing ensemble averages is a fundamental procedure in the statistical analysis of dynamical models. When an external field is applied to the system of interest, its response is determined by the average of many individual ensemble members evolving in the forced scenario. The quality of this averaging procedure, however, heavily depends on the nature of the underlying system. The strength of chaos through Lyapunov exponents \cite{Eckmann1985}, the mixing behaviour of the system \cite{lasota}, the nature of the observation functions \cite{lasota} and the dimensionality of the system \cite{milinski2020} are some factors that heavily influence the computations of ensemble averages.


The transient-time correlation function (TTCF) method \cite{evans_morris_1990,evans_searles_2016} is widely used in non-equilibrium fluids to compute transport coefficients in turbulent flows, molecular systems \cite{morris_1987,smith2015,MAFFIOLI2024109205} and other many-body systems \cite{carbone_2026}. These transport phenomena arise upon the introduction of a non-conservative perturbation force into a thermostatted system with statistics initially obeying the canonical ensemble. To compute non-equilibrium statistical quantities, the thermostatted system at equilibrium is perturbed and time-dependent ensemble averages are calculated. Obtaining such average might seem straightforward, but due to the large number of interacting particles in the system, the signal-to-noise ratio (SNR) can be extremely poor. For this, the TTCF method, which allows one to write the response of the system in terms of suitably defined correlations, provides a way out, based on nonperturbative response theory, to improve the quality of ensemble averages, without resorting to a prohibitive amount of ensemble members to accurately sample the reference stationary state. 

The TTCF method still lacks a systematic and analytic assessment, especially in cases where the reference system is far from thermodynamic equilibrium (i.e. it does not obey detailed balance) and a further finite-size perturbation is applied. 
Although the notion of a dissipation function, which is  central to TTCF, has been examined in the stochastic setting \cite{colangeli_2025}, several issues arise in this wider context. The main problem is that most non-equilibrium systems encountered in applications typically lack a closed-form invariant measure, making it difficult to characterize their stationary statistics \cite{majda_abramov_book}. Hence, one needs to resort to approximations such as those based on Gaussian mixtures \cite{abramov2007,cooper_haynes_2011} or to take advantage of generative modelling \cite{giorgini2024linear,giorgini2024datadriven,giorgini2025statistical}.



In the limit case where a weak forcing is applied to the system, TTCF is closely related to the fluctuation-dissipation theorem (FDT) \cite{kubo}, which provides a link between natural and forced fluctuations of a system, or, more precisely, allows one to express any linear response operators as a correlation function of the unperturbed system. The FDT has been extended to nonequilibrium systems possessing an invariant measure that is absolutely continuous with respect to Lebesgue \cite{marconi2008,Hairer2010,Sarracino2019}, as in the case of elliptic diffusion processes \cite{pavliotisbook2014}, whilst the theorem does not apply in the case of dissipative chaotic systems \cite{ruellegeneral1998,ruelle2009}.
Linear response for diffusion processes has  been developed in functional analytical terms by taking advantage of the Koopman operator theory \cite{Mezic2005,Koopman,KoopmanSIAMReview}, linking it with the spectral features of the system in responding to external perturbations \cite{Santosgutierrez_2022,LucariniChekroun2023,LucariniChekroun2024}. On a closely related angle, in a series of papers \cite{Lucarini2016,SantosJSP,lucarini2025interpretableequationfreeresponsetheory} we have shown that Markov chains provide an extremely favourable environment for rigorously deriving response formulas that can  be readily used for studying the response of a large class of systems. 

It is then attractive to re-examine TTCF in light of these recent theoretical developments with the goal of clarifying its applicability in a vast range of systems.  
The aim of this work is to first derive TTCF in the context of  diffusion processes and of Markov chains, which allow us to cover the dynamics of a general system whose unperturbed state is far from equilibrium. We then examine the performance of the TTCF method in a class of prototypical stochastic dynamical systems outside the traditional near-equilibrium setting, with a specific emphasis on scenarios where the base system does not obey detailed balance and probability fluxes are nontrivial. We will systematically study in the case of a two-dimensional Ornstein-Uhlenbeck process how bringing the system far from equilibrium changes the performance of TTCF. We will also analyze a the stochastically perturbed version of the Lorenz'96 model, which is one of the most relevant conceptual models in geophysical fluid dynamics  \cite{lorenz1996predictability,lorenz1998optimal,wilks_effects_2005,lucarini2011statistical,gallavotti2014equivalence} .  



The paper is structured as follows. In Sect. \ref{sec: nonlinear fluctuation response relation} we provide a derivation of TTCF in the broad context of diffusion processes. In Sect. \ref{signatonoise} we evaluate the performance of the TTCF in terms of SNR. In Sect. \ref{numerical} we present results obtained on different numerical models of different level of complexity. In Sect. \ref{conclusions} we present our conclusions. The paper is complemented by App.~\ref{app:derivation_variance}, where we derive the variance of the TTCF estimator, App. \ref{rotation}, where we discuss how the performance of TTCF changes as we go from equilibrium to far from equilibrium conditions in linear processes, and App. \ref{TTCFMarkov}, where we derive TTCF using the formalism of Markov chains.

\section{Derivation of the nonlinear fluctuation-response relation}
\label{sec: nonlinear fluctuation response relation}
The first derivations of the relation were done for thermostatted molecular systems \cite{morris_1987,evans_morris_1990,evans_searles_2016}.  Here, we take the stochastic differential equation (SDE) approach to adapt the formulas to systems possessing variables evolving on fast timescales modeled by noise. The derivations done in \cite{Iannella2023-tb,colangeli_2025} provide a definition of the dissipation function for stochastic systems and non-perturbative response formulas. In this section we revisit such derivation.

Let us consider an elliptic diffusion process associated with a $d$-dimensional It\^o SDE with deterministic drift $F : \mathbb{R}^d \longrightarrow \mathbb{R}^d$, perturbation field $G:\mathbb{R}^d \longrightarrow \mathbb{R}^d$ and stochastic part given by the $d\times p$ matrix $\Sigma \in \mathbb{R}^{d\times d}$, $p\geq 1$:
\begin{equation}
 \label{eq:sto ode 2}
\mathrm{d} X_t  = \left[F (X_t)  + \varepsilon G(X_t) \right]\mathrm{d} t+\Sigma(X_t)\mathrm{d} W_{t},
\end{equation}
where $W_t$ is a $d$-dimensional Wiener process and $\varepsilon>0$ and the correlation matrix $\Sigma(X_t)\Sigma(X_t)^\top$ is nowhere degenerate.  
The evolution of probability density functions associated with the SDE \eqref{eq:sto ode 2}, is given by the Fokker-Planck equation \cite{risken}:
\begin{equation}\label{eq:fpe 2}
\partial _t\rho = \left( \hat{\mathcal{L}}_0 + \varepsilon \hat{\mathcal{L}}_1 \right)\rho= \underbrace{-\nabla \cdot \left( F\rho \right)   +\frac{1}{2}\nabla^2:\left(  \Sigma \Sigma^{\top}\rho \right)}_{\hat{\mathcal{L}_0}} - \underbrace{\varepsilon  \nabla \cdot \left( G\rho \right)}_{\hat{\mathcal{L}}_1},
\end{equation}
where the symbol ``$:$'' denotes the entry-wise product and $\hat{\mathcal{L}}_0$ and $\hat{\mathcal{L}}_1$ are linear operators acting on the same Hilbert space. We define $\rho_0$ as the density associated with the invariant measure of the unperturbed system, such that  
\begin{equation}\label{eq:rho0}
\partial _t\rho_0 = \hat{\mathcal{L}}_0\rho_0=0
\end{equation}
In what follows, we will consider initial conditions $X_0 \sim \rho_0$. The Fokker-Planck approach corresponds to a perspective in which the evolution of densities, that is, loosely speaking, volumes of initial points in the phase space, is tracked as time evolves. Usually, the \textit{adjoint semigroup} $\hat{P}_t^\varepsilon : L^1 \to L^1 $ acting on integrable densities $\rho : L^1 \to \mathbb{R}$ is introduced as
\begin{equation}
    (\hat{P}_t^\varepsilon \rho ) (y) = \int p_\varepsilon(y,t|x,0)\rho(x)\mathrm{d}x,
\end{equation}
where $p_\varepsilon(y,t|x,0)$ is the transition density originated by the perturbed flow, satisfying the Fokker-Planck equation \eqref{eq:fpe 2}. The adjoint semigroup pushes densities forward in time and can be written as $\hat{P}_t^\varepsilon = e^{t(\hat{\mathcal{L}}_0 + \varepsilon\hat{\mathcal{L}}_1)}$. We are here interested in understanding how perturbations, encoded in the perturbation field $G$ and activated when $\varepsilon\neq 0$, act on the invariant distribution of the unperturbed system $\rho_0$ which solves $\hat{P}_t^{0}\rho_0 = \rho_0$, or, equivalently, $\hat{\mathcal{L}}_0\rho_0 =0$. In other words, we are interested in evaluating the time evolution $\rho(t,\cdot) = \hat{P}^\varepsilon_t\rho_0$.

One can also adopt an alternative picture that focuses on evolution of observables, which are measurable functions of the phase space. This picture is provided by the Markov semigroup associated to the stochastic process \eqref{eq:sto ode 2}. Given an observable $\Psi :\mathbb{R}^d \to \mathbb{R} $, we can define the Markov semigroup $P_t^\varepsilon$ as
\begin{equation}
    (P^\varepsilon_t\Psi)(x) = \mathbb{E}_x[\Psi(X_t)] = \int \Psi(y)p_\varepsilon(y,t|x,0)\mathrm{d}y.
\end{equation}
 In the previous equation we have used the notation for the conditional expectation value $\mathbb{E}_x[\Psi(X_t)] = \mathbb{E}[\Psi(X_t)|X_0=x] $. For a fixed initial condition,  the Markov semigroup determines the evolution of the statistical properties, averaged over all noise realisations. It is natural to consider the Markov semigroup as acting on observables $\Psi$ that belong to $L^2(\rho_0)$, the space of square integrable functions with respect to the invariant measure of the unperturbed system. The two approaches are equivalent,  with the adjoint semigroup and the Markov semigroup being dual to each other with the canonical $L^2$ scalar product, $\langle P_t^\varepsilon \Psi , \rho\rangle = \langle \Psi, \hat{P}_t^\varepsilon \rho \rangle$.  
 
As mentioned before, we are interested in evaluating the response of the system to the perturbation field $G$. Therefore, we are interested in evaluating the expectation value of observables 
\begin{equation}
\label{eq: response}
    \langle \Psi \rangle_\varepsilon(t) \coloneqq \mathbb{E}[\Psi(X_t)] = \mathbb{E}[\mathbb{E}_x[\Psi(X_t)]] = \int (P^\varepsilon_t\Psi)(x)\rho_0(x)\mathrm{d}x,
\end{equation}
where in the second equality we have used the tower rule, that is we first evaluate the action of the Markov semigroup given an initial condition and then we average over all initial conditions. 
We can find a nonlinear response formula, at the basis of the TTCF approach, by evaluating the time derivative of the response

\begin{equation}
    \frac{\mathrm{d}\left\langle \Psi \right\rangle_{\varepsilon}}{\mathrm{d}t} = \frac{\mathrm{d}}{\mathrm{d}t}\int \Psi(x)\hat{P}^\varepsilon_t\rho_0 \mathrm{d}x= \int \Psi(x) \hat{P}^\varepsilon_t \left(\hat{\mathcal{L}}_0 + \varepsilon \hat{\mathcal{L}}_1\right)\rho_0 \mathrm{d}x = \varepsilon \int (P_t^\varepsilon\Psi)(x)\hat{\mathcal{L}}_1\rho_0 \mathrm{d}x,
\end{equation}
where we have used the duality pairing of the semigroups, the exponential form of the adjoint semigroup and the fact that $\hat{\mathcal{L}}_0\rho_0=0$. 
We now introduce the score or dissipation function $\Omega (\mathbf{x}) = \mathcal{L}_{1}\rho_0(\mathbf{x}) / \rho_0(\mathbf{x})$, to rewrite the previous expression in form of a lagged correlation function
\begin{equation}\label{eq:derivative_epsilon}
        \frac{\mathrm{d}\left\langle \Psi \right\rangle_{\varepsilon}}{\mathrm{d}t} =\varepsilon\int (P_t^\varepsilon\Psi)(x)\Omega(x)\rho_0(x)\mathrm{d}x=\varepsilon \mathbb{E}\left[ (P_t^\varepsilon\Psi)(X) \Omega(X)\right] ,
\end{equation}
Integrating in time we arrive at the following expression nonlinear response formula \cite{morris_1987,evans_searles_2016,colangeli_2025}:
\begin{equation}\label{eq:ttcf}
    \left\langle \Psi \right\rangle_{\varepsilon}(t) -\left\langle \Psi \right\rangle_{0} = \varepsilon \int_0^t \mathbb{E}\left[\left(P^{\varepsilon}_s\Psi\right)(X)\Omega(X)\right]\mathrm{d}s.
\end{equation}
This is a nonperturbative result which relates ensemble averages computed $t$ time units after the perturbation is applied with a correlation function taken during such interval of time of the observable $\Psi$ and $\Omega$, where the expectation is taken with respect to $\rho_0$, yet propagated by the perturbed flow. We remark that this is an exact result valid for all values of $\varepsilon$; see App. \ref{TTCFMarkov} for an equivalent result obtained for Markov chains. If $\varepsilon$ is assumed to be small, the above result yields the classical fluctuation dissipation theorem as we will see below.

\section{SNR for response estimators: direct averages vs nonlinear fluctuation-response}\label{signatonoise}
Equation~\eqref{eq:ttcf} relates two theoretically identical quantities, although different in practical terms, as discussed in \cite{Todd_Daivis_2017,MAFFIOLI2024109205}. The difference strikes in that taking direct averages in chaotic, stochastic and non-equilibrium systems can be an extremely ill-posed problem and, in such regard, the TTCF method provides better SNRs in computing the same quantities \cite{MAFFIOLI2024109205}. To gain analytical insight, we seek a Monte Carlo estimation of both sides of the response equation. We consider a finite set $\{X_i\}_{i=1}^N$ of independent and identically distributed variables, where $X_i \sim \rho_0$ for all $ i$. The two sides of equation \eqref{eq:ttcf} are estimated respectively as   
\begin{subequations} \label{eq:estimators_R_t_C_t}
\begin{align}
R_t &= \frac{1}{N}\sum_{i=1}^N\Psi(X^{(i)}_t) - \langle \Psi \rangle_0 \\
C_t &= \frac{\varepsilon}{N}\sum_{i=1}^N\int_0^t\Psi(X_s^{(i)})\Omega(X_i)\mathrm{d}s,
\end{align}
\end{subequations}
where $X_t^{(i)}$ is the stochastic process originated from Eq.~$\eqref{eq:sto ode 2} $ with initial condition $X_0^{(i)} = X_i$ and independent noise realization (Wiener Process). Note that we consider $\langle \Psi \rangle_0$ as a known parameter; it is not part of the estimation procedure. Operationally, one could think that very long trajectories of the unperturbed system are available so that the unperturbed average of the observable is known with very good accuracy. $R_t$ corresponds to a direct averages estimator for the response, whereas $C_t$ is the estimator from the TTCF method. Here, we want to investigate the SNR of the estimators. For a given estimator, say $R_t$, the SNR is defined as 
\begin{equation}\label{eq:SNR_def}
    \mathrm{SNR}[R_t] = \frac{|\mathbb{E}\left[ R_t \right]|}{\sqrt{\mathbb{E}\left[ R_t^2 \right] - \mathbb{E}\left[R_t\right]^2}}.
\end{equation}
An equivalent definition holds for $C_t$. As introduced in section \ref{sec: nonlinear fluctuation response relation}, the expectation values are taken over the initial distribution $\rho_0$ and, fixed an initial condition, over all noisy paths. 
\paragraph{Unbiasedness of the estimators}
It is simple to see that both Monte Carlo estimators are unbiased:
\begin{equation}
    \mathbb{E}[R_t] = \frac{1}{N}\sum_i^N\mathbb{E}[\Psi(X_t^{(i)})] - \langle \Psi \rangle_0 = \mathbb{E}[\Psi(X_t)]  - \langle \Psi \rangle_0,
\end{equation}
where the last equality comes from the fact that $X_i$ are identical random variables and $X_t$ is a stochastic process originated by \eqref{eq:sto ode 2} with initial condition $X_0 = X \sim \rho_0$. This is exactly the definition of the response $\langle \Psi\rangle_\varepsilon - \langle \Psi \rangle_0$ in \eqref{eq:ttcf}.
Similarly, the expectation value of the TTCF estimator is 
\begin{equation}\label{eq:mean_c_t}
    \mathbb{E}[C_t] = \varepsilon \int_0^t\mathbb{E}[\Psi(X_s)\Omega(X)]\mathrm{d}s.
\end{equation}
Now, this is exactly the right hand side of \eqref{eq: response}, since, because the law of total expectation \cite{billingsley1995probability}, we have that
\begin{align}
    \mathbb{E}[\Psi(X_s)\Omega(X)] = \mathbb{E}[ \mathbb{E}_x[\Psi(X_s)]\Omega(X)] = \mathbb{E}[(P^\varepsilon_s\Psi)(X)\Omega(X)] .
\end{align}
For notational convenience, we define $\Gamma^\varepsilon (s) =\mathbb{E}\left[(P^{\varepsilon}_s\Psi)(X)\Omega(X)\right]$.

\paragraph{Variance and signal to noise ratio of the estimators}
Since $\{X_i\}_{i=1}^N$ are identically distributed and independent random variables, and since we consider independent realisations of the random paths for each $i$, we have that  $\mathbb{E}[f(X_s^{(i)})f(X_t^{(j)})]= \mathbb{E}[f(X_s^{(i)})]\mathbb{E}[f(X_t^{(j)}]  $ for $i\neq j$, any deterministic function $f$ of the random variable and for all $ t,s >0$. Using the previous independence property for $f=\Psi$ gives the variance of the estimator $R_t$ as
\begin{equation}
 \mathrm{Var}[R_t] =  \mathbb{E}[R_t^2] - \mathbb{E}[R_t]^2 = \frac{1}{N}\left( \langle \Psi^2 \rangle_\varepsilon- \langle \Psi \rangle^2_\varepsilon\right).
\end{equation}
The SNR of the direct averages estimator is
\begin{equation}
    \mathrm{SNR}[R_t ]  = \frac{|\mathbb{E }[R_t]|}{\sqrt{\mathrm{Var}[R_t]}}=\sqrt{N}\frac{| \langle \Psi \rangle_\varepsilon  - \langle \Psi \rangle_0|}{\sqrt{\langle \Psi^2 \rangle_\varepsilon- \langle \Psi \rangle^2_\varepsilon}}.
\end{equation}
As expected, the variance decreases as more ensemble members are considered, and depends on the instantaneous variance of the observable $\Psi$.



A similar, but lengthier calculation allows us to find the second moment of the TTCF estimator:
\begin{equation}\label{eq:snr_ttcf}
    \mathrm{SNR}[C_t] = \sqrt{N}\frac{|\int_{0}^t \Gamma^\varepsilon (s)\mathrm{d}s|}{\sqrt{\int_{0}^t\int_0^t \bigg(\mathbb{E}\left[(P^{\varepsilon}_s\Psi)(x)\Omega(x)(P^{\varepsilon}_u\Psi)(x)\Omega(x) \right] - \Gamma^\varepsilon(s)\Gamma^\varepsilon(u) \bigg)\mathrm{d}s\mathrm{d}u}}.
\end{equation}
Again, the SNR improves, slowly, as more ensemble members $N$ are considered. In the next section we will investigate the scaling of the SNR with respect to the amplitude of the forcing $\varepsilon$.

\subsection{Scaling of SNR for weak forcings}
In this section we show that for small forcing, the SNR of direct averages scales as $\mathcal{O}(\varepsilon)$ whereas the TTCF method scales as $\mathcal{O}(1)$. The results of this section provides a theoretical justification of what has been suggested previously in \cite{MAFFIOLI2024109205}. We consider a perturbation expansion of the Markov semigroup associated with the perturbed flow
\begin{equation}
    P_t^\varepsilon = P_t^{(0)} + \varepsilon P_t^{(1)} + \mathcal{O}(\varepsilon^2).
\end{equation} 
In this perturbative approach, the response of the observable $\Psi$ is given by $\langle \Psi\rangle_\varepsilon = \langle \Psi\rangle_0 + \varepsilon \langle \Psi \rangle_1 + \mathcal{O}(\varepsilon^2)$ where $\langle \Psi \rangle_1 = \int (P_t^{(1)}\Psi)(x)\rho_0(x)\mathrm{d}x$ is the linear response of the system. 
The SNR of the direct averages estimator is 
\begin{equation}\label{eq:da_limit_epsilon}
        \mathrm{SNR}[R_t] = \varepsilon \sqrt{N}  \frac{|\langle \Psi \rangle_1(t) |}{\sqrt{\langle \Psi^2\rangle_0 - \langle \Psi \rangle_0^2  } }+\mathcal{O}(\varepsilon^2) .
\end{equation}
This leading order term depends on the depends on the linear sensitivities of the system with respect to the amplitude of the perturbation $\varepsilon$.

We now evaluate the SNR for the TTCF estimators for weak forcings. We first observe that in a weak forcing approximation, the nonlinear response formula \eqref{eq:ttcf} corresponds to the usual fluctuation-dissipation theorem. Using the expansion of the semigroup in \eqref{eq:ttcf} we have that
\begin{equation}
    \varepsilon \langle \Psi \rangle_1 + \mathcal{O}(\varepsilon^2) = \varepsilon \int_0^t \mathbb{E}[(P_s^{(0)}\Psi)(x)\Omega(x)]\mathrm{d}s + \mathcal{O}(\varepsilon^2) = \varepsilon \int_0^t \Gamma^{(0)}(s) + \mathcal{O}(\varepsilon^2)
\mathrm{d}s,
\end{equation}
where $\Gamma^\varepsilon(t)= \Gamma^{(0)}(t) + \mathcal{O}(\varepsilon)$ is the perturbation expansion of $\Gamma$ due to the perturbation expansion of the Markov semigroup. The numerator of SNR for the TTCF estimator is thus $|\langle \Psi \rangle_1 | + \mathcal{O}(\varepsilon)$.
On the other hand, it is not clear how the denominator of Eq.~\eqref{eq:snr_ttcf} scales with respect to $\varepsilon$. If one replaces the time-evolution in Eq.~\eqref{eq:var_c} by the unperturbed dynamics and, without loss of generality, assumes that the unperturbed dynamics evolve at a steady-state, one can write the integrand in Eq.~\eqref{eq:snr_ttcf} as:
\begin{equation}
    \mathbb{E}\left[ (P_s^\varepsilon\Psi)(x)\Omega(\mathbf{x})(P_u^\varepsilon\Psi)(x)\Omega(\mathbf{x}) \right] - \Gamma(s)\Gamma(u) = \chi_0(s,u) + \varepsilon\chi_1(s,u) + \mathcal{O}(\varepsilon^2),
\end{equation}
where $\chi_0(\cdot,\cdot)$ is a zeroth-order term representing a four-point correlation function; see also the derivation of perturbation expansions of correlation functions in \cite{Lucarini_wouters_2017}. Hence, in the limit of small $\varepsilon$:
\begin{equation}\label{eq:ttcf_limit_epsilon}
    \mathrm{SNR}(C_t) = \sqrt{N}\frac{|\langle \Psi \rangle_1(t)|}{\sqrt{\int_{0}^t\int_{0}^t\chi_0(s,u)\mathrm{d}s\mathrm{d}u}} + \mathcal{O}(\varepsilon).
\end{equation}

Gathering the results of Eq.~\eqref{eq:da_limit_epsilon} and \eqref{eq:ttcf}, we conclude that the SNR of taking direct averages is worse for weak forcings, compared to the TTCF method. In fact, in the limit of $\varepsilon\rightarrow 0$, the SNR of taking direct averages vanishes, whereas for the TTCF it stays positive.

\subsection{The role of the spectrum of the generator: SNR time-evolution}

The coefficient multiplying $\sqrt{N}$ in Eq.~\eqref{eq:ttcf_limit_epsilon} is a function of time. It is worth examining the time-dependence of this coefficient. Let us define such coefficient $h(t)$ as:
\begin{equation}\label{eq:eigendecomp}
    h(t) = \frac{|\langle \Psi \rangle_1(t)|}{\sqrt{\int_{0}^t\int_{0}^t\chi_0(s,u)\mathrm{d}s\mathrm{d}u}}.
\end{equation}
The idea now is to employ the spectral decomposition of correlation and response function in terms of the eigenvalues and eigenfunctions of the Fokker-Planck operator $\hat{\mathcal{L}}_{0}$ defined in Eq.~\eqref{eq:fpe 2}. Assuming that there are $J$ main contributing eigenvalues and eigenfunctions, we can expand the linear response function in the  the numerator of $h(t)$ as \cite{chekroun2019c,Santosgutierrez_2022}:
\begin{equation}
    \langle \Psi \rangle_1(t) = \int_0^t \sum_{j=1}^J\alpha_j e^{\lambda_js}\mathrm{d}s = \sum_{j=1}^J\frac{\alpha_j}{\lambda_j}\left(e^{\lambda_jt} - 1 \right).
\end{equation}
Since we are considering all conjugate pairs, the sum is a real number in spite of $\lambda_j$ being a complex number. The values of $\alpha_j$ can be found precisely as the projection of the observables $\Psi$ and $\Omega$ onto the eigenfunctions of $\mathcal{L}_0$ \cite{Santosgutierrez_2022}. We shall define $\lambda_1$ to be the smallest eigenvalue (in absolute value) different from zero. This eigenvalue gauges the characteristic time for the system governed by $\mathcal{L}_0$ to converge to steady-state.

We seek a spectral decomposition of $\chi_0$, which precisely reads as:
\begin{subequations}
\begin{align}
    \chi_0(s,u) &= \mathbb{E}\left[ (P_u\Psi)(x)\Omega(x)(P_s\Psi)(x)\Omega(x) \right] - \Gamma(s)\Gamma(u)\\&= \mathbb{E}\left[\Omega(x)^2 P_u(\Psi P_{s-u}\Psi)(x) \right] - \Gamma(s)\Gamma(u).
\end{align}
\end{subequations}
Let us assume that $u\leq s$. Consider the following spectral decompositions in terms of the $J$ eigenfunctions of the semigroup:
\begin{align}
    \Psi &= \sum_{j=0}^J a_j\varphi_j,\quad \text{with} \quad a_j = \langle \varphi^{\ast}_j,\Psi \rangle \\
    \Omega^2 &= \sum_{j=0}^J b_j\varphi^\ast_j,\quad \text{with} \quad b_j = \langle \Omega^2 ,\varphi_j\rangle \\
\Psi P_{s-u}\Psi &= \sum_{j=0}^Jc_j(s-u)\varphi_j \quad \text{with} \quad c_j(s-u)
= \sum_{\ell=0}^{J}\sum_{k=0}^{J}
a_\ell a_k e^{\lambda_k (s-u)}
\left\langle \varphi_j^{\ast}, \varphi_\ell \varphi_k \right\rangle,
\end{align}
where we have used that $\Psi P_{s-u}\Psi = \sum_{j,k}a_ja_ke^{\lambda_k(s-u)}\varphi_j\varphi_k$. We now take the expectation values $\mathbb{E}\left[\cdot\right]$ to obtain:
\begin{equation}\label{eq:chi_0_spectral}
    \chi_0(s,u) = \sum_{j=0}^J \sum_{\ell=0}^{J}\sum_{k=0}^{J}
T_{j\ell k} e^{\lambda_ju}e^{\lambda_k (s-u)} - \sum_{j,k=1}^J A_{jk}e^{\lambda_js+\lambda_ku},
\end{equation}
where $T_{j\ell k} = b_ja_\ell a_k\left\langle \varphi_j^{\ast}, \varphi_\ell \varphi_k \right\rangle$ and $A_{jk}=\alpha_j\alpha_k$. Because $\Omega(\cdot)^2$ is a strictly positive continuous function, it follows that its mean with respect to $\rho_0$ is non-zero. Hence, $b_0\neq 0$ in Eq.~\eqref{eq:chi_0_spectral}. This means that $\chi_0$ can be split into two so that $\chi_0(s,u) = \chi_0^{+}(s-u) + \chi_0^{-}(s,u)$ with:
\begin{subequations}
\begin{align}
        \chi_0^{+}(s-u) &= \sum_{\ell=0}^{J}\sum_{k=0}^{J}
T_{0\ell k} e^{\lambda_k (s-u)}
 \\
\chi_0^{-}(s,u)  &= \sum_{j=1}^J \sum_{\ell=0}^{J}\sum_{k=0}^{J}
T_{0\ell k} e^{\lambda_ju}e^{\lambda_k (s-u)}
- \sum_{j,k=1}^J A_{jk}e^{\lambda_js+\lambda_ku} .
\end{align}
\end{subequations}
Now we need to extend it for all times $s$ and $u$. Because we are assuming that the reference state is stationary, the function $\chi_0(\cdot,\cdot)$ satisfies symmetry in both arguments. In particular, the function $\chi_0^{+}$ is written as:
\begin{equation}
    \chi^{+}_0(s-u) = \sum_{\ell=0}^{J}\sum_{k=0}^{J}
T_{0\ell k} e^{\lambda_k |s-u|}.
\end{equation}
Moreover, because $\mathrm{Re}(\lambda_j)<0$ for $j=1,\ldots,J$, there exist two constants $\gamma, Z > 0$ such that:
\begin{equation}
     |\chi^{-}_0(s,u)|\leq Z\left( e^{-\gamma u} + e^{-\gamma(u+s)} \right).
\end{equation}
Hence, the square of the denominator of $h(t)$ is:
\begin{subequations}
\begin{align}
\int_{0}^t\int_0^t \chi_0(s-u)\mathrm{d}s\mathrm{d}u &= \int_{0}^t\int_0^t \chi_0^{+}(s-u)\mathrm{d}s\mathrm{d}u + \int_{0}^t\int_0^t \chi^-_0(s,u)\mathrm{d}s\mathrm{d}u \\ &= \sum_{\ell,k=1}^J\frac{2T_{0\ell k}}{\lambda_k ^2}\left( e^{t\lambda_k}-1- \lambda_k t\right) + \int_{0}^t\int_0^t \chi^-_0(s,u)\mathrm{d}s\mathrm{d}u.
\end{align}
\end{subequations}

For initial times $t\ll 1/|\lambda_J|$, we can expand the exponentials in the small variable $t$ to obtain the following expression:
\begin{equation}
    h(t)= 	  \frac{\sum_{j=1}^J\frac{\alpha_j}{\lambda_j}\left(\lambda_jt + \mathcal{O}(t^2) \right)}{\sqrt{\sum_{\ell,k=1}^J \frac{2T_{0\ell k}}{\lambda_k ^2}\left( \frac{\lambda_k^2}{2}t^2\right)+\chi_0^-(0,0)t^2}} = \frac{\sum_{j=1}^J\alpha_j}{\sqrt{\sum_{j=1}^JT_{0\ell k}+\chi_0^-(0,0)}} +\mathcal{O}(t^2).
\end{equation}
The value in the numerator is nothing other than the initial value of the response function, which is equal to the covariance between $\Psi$ and $\Omega$ at time $t=0$ with respect to the stationary state $\rho_0$.

For times larger than the characteristic time of the unforced system, namely, $t\gg 1/|\lambda_1|$, the exponentials converge to zero, so that we arrive at the following expression:
\begin{equation}\label{eq:h_spectrum}
    h(t) = \frac{\sum_{j=1}^J\frac{\alpha_j}{\lambda_j}}{\sqrt{2t\sum_{\ell,k=1}^J-\frac{T_{0\ell k}}{\lambda_j}+I}},
\end{equation}
where $I$ is defined as the large-time limit of the double integral of $\chi_0^{-}(s,u)$ and satisfies:
\begin{subequations}
\begin{align}
    |I| &= \left| \lim_{t\rightarrow \infty}\int_0^t\int_0^t \chi_0^-(s,u)\mathrm{d}u\mathrm{d}s \right| \leq \lim_{t\rightarrow \infty}\int_0^t\int_0^t |\chi^-_0(s,u)|\mathrm{d}u\mathrm{d}s \\ &\leq \lim_{t\rightarrow \infty}\int_0^t\int_0^t  Z\left( e^{-\gamma u} + e^{-\gamma(u+s)} \right)\mathrm{d}u\mathrm{d}s <\infty,
\end{align}
\end{subequations}
since $\gamma$ is a strictly positive constant.

Equation~\eqref{eq:h_spectrum} tells us that as time progresses, the SNR of the TTCF method becomes smaller with time, decreasing with the square root of time for large times. The actual rate depends explictly on the spectral coefficients. Particularly, the relation in Eq.~\eqref{eq:h_spectrum}, reveals that the projection coefficients $\alpha_j$ and $T_{0\ell k}$ determine the signal retrieved. Indeed, if an observable $\Psi$ projects entirely around, say, the eigenvalue of smallest real part $\lambda_1$, the numerator in Eq.~\eqref{eq:h_spectrum} will be large and $\alpha_j=0$, for $j\neq 1$. The denominator, on the other hand, relates to the projection of $\Psi$ and $\Omega$ through the correlation functions. This means that more eigenvalues will be excited and the weights $|T_{0\ell k}|<|\alpha_{1}|$, for any $j=1,\ldots,J$.

Repeating the same analysis for $\mathrm{SNR}(D(t))$ yields limiting behaviour for values of long and short times. It suffices to replace in Eq.\eqref{eq:da_limit_epsilon} the eigen-decomposition shown in Eq.~\eqref{eq:eigendecomp} and to realize that the variance of the observable $\Psi$ has a non-trivial zero-order term in $\varepsilon$. Hence, for $t\ll -1/\mathrm{Re}(\lambda_J)$, we have that
\begin{equation}
    \mathrm{SNR}(D(t))  =  \varepsilon \sqrt{N} t \frac{\sum_{j=1}^J\alpha_j}{\sqrt{\left\langle \Psi^{2} \right\rangle_{0}}} + \mathcal{O}(\varepsilon^2),
\end{equation}
which reveals a linear time dependence and a particularly bad SNR at early times. Contrarily, for large values of time, $t\gg -1/\mathrm{Re}(\lambda_1)$, we have that the SNR saturates (to leading order in $\varepsilon$):
\begin{equation}
    \mathrm{SNR}(D(t))=-\varepsilon \sqrt{N} \frac{\sum_{j=1}^J\frac{\alpha_j}{\lambda_j}}{\sqrt{\left\langle \Psi^{2} \right\rangle_{0}}} + \mathcal{O}(\varepsilon^2).
\end{equation}
The value of the saturation of $\mathrm{SNR(D(t))}$ depends on two quantities, apart from $\varepsilon$ and $N$. It depends on the spectral coefficients $\alpha_j/\lambda_j$ and the variance of $\Psi$ in the unperturbed system. While $\sqrt{\langle\Psi^2\rangle_0}$ does not depend on the spectrum of $\mathcal{L}_0$, the magnitude of $\alpha_j/\lambda_j$ depends on how the linear response function projects in eigenvalues with small or large real parts. For instance, projecting in faster modes yields larger values of $\mathrm{Re}(\lambda_j)$.

\section{Numerical examples}\label{numerical}
\subsection{A one-dimensional OU process}
In this section we investigate the one-dimensional Ornstein-Uhlenbeck (OU) process. Here we address two aspects: 1) What is the validity of TTCF in stochastic systems and, 2) the role of the spectral decomposition of observable function in computing its average after applying a forcing. We thus consider the following stochastic differential equation (SDE) for the real variable $x(t)$:
\begin{equation}\label{eq:ou1d}
    \mathrm{d}x(t)= ax(t)\mathrm{d}t + \varepsilon\mathrm{d}t + \sigma \mathrm{d}W_{t},
\end{equation}
where $a<0$, $\varepsilon,\sigma>0$ and $W_t$ is a one-dimensional Wiener process. In this case, we apply a constant perturbation with strength $\varepsilon$.

To estimate the response we aim at computing the direct averages--- on the left-hand side of Eq.~\eqref{eq:ttcf}--- and the TTCF method--- on the right-hand side of Eq.~\eqref{eq:ttcf}. For this we need the value of the observable $\Omega$. Knowing that the invariant distribution of Eq.~\eqref{eq:ou1d} for $\varepsilon=0$ is $\rho_0(x) = \sqrt{\frac{-a}{\pi \sigma^2}}e^{2ax^2/\sigma^2 }$, we just apply the definition for $\Omega$:
\begin{equation}\label{eq:Omega_1d_ou}
    \Omega(x) = -\frac{\partial_x\left(\varepsilon\rho_0(x)\right)}{\rho_0(x)} = -\varepsilon \partial_x \log \left(e^{ax^2/\sigma^2} \right) = -\frac{2\varepsilon a}{\sigma^2}x.
\end{equation}
This is enough to apply the TTCF method.


The Fokker-Planck operator $\mathcal{L}_0$--- from Eq.~\eqref{eq:fpe 2}--- associated with Eq.~\eqref{eq:ou1d} has well known analytical features \cite{metafunes2002,pavliotisbook2014}. In particular, the spectrum of this operator is a pure-simple-point spectrum give by $\{an\}_{n=0}^\infty$. Furthermore, the associated left-eigenfunctions $\{\varphi_n\}_{n=1}^\infty$ are determined by Hermite polynomials $\mathrm{He}_n(x)$:
\begin{equation}\label{eq:hermite}
    \varphi_n(x)
= \mathrm{He}_n\!\left( -\frac{\sqrt{-2a}}{\sigma}\, x \right),
\end{equation}
for $n=0,1,2,\ldots$ This gives a simple framework to assess the role of the spectrum in the properties of convergence of direct averages and the TTCF method.

 By examining the eigenfunctions given in Eq.~\eqref{eq:hermite}, it is trivial to see that $\Psi(x)=x$ projects entirely onto $\mathrm{He}_1\!\left( -\frac{\sqrt{2a}}{\sigma}\, x \right)$. We investigate three observables $\Psi_i$ defined by\begin{equation}
	\Psi_i(x) = \left(\frac{\sqrt{-2a}}{\sigma}x\right)^{1+2^{(i-1)}},
\end{equation}
for $i=1,2,3$. While $\Psi_1$ is proportional to the observable studied earlier, $\Psi_2$ and $\Psi_3$ project respectively on nontrivial components relative to  $\mathrm{He}_3$ and $\mathrm{He}_5$, respectively. This means that they have a nontrivial projection onto the eigenvalues $3a$ and $5a$, of the operator $\mathcal{L}_0$.

\begin{figure}[H]
	\centering
	\includegraphics[scale=0.3]{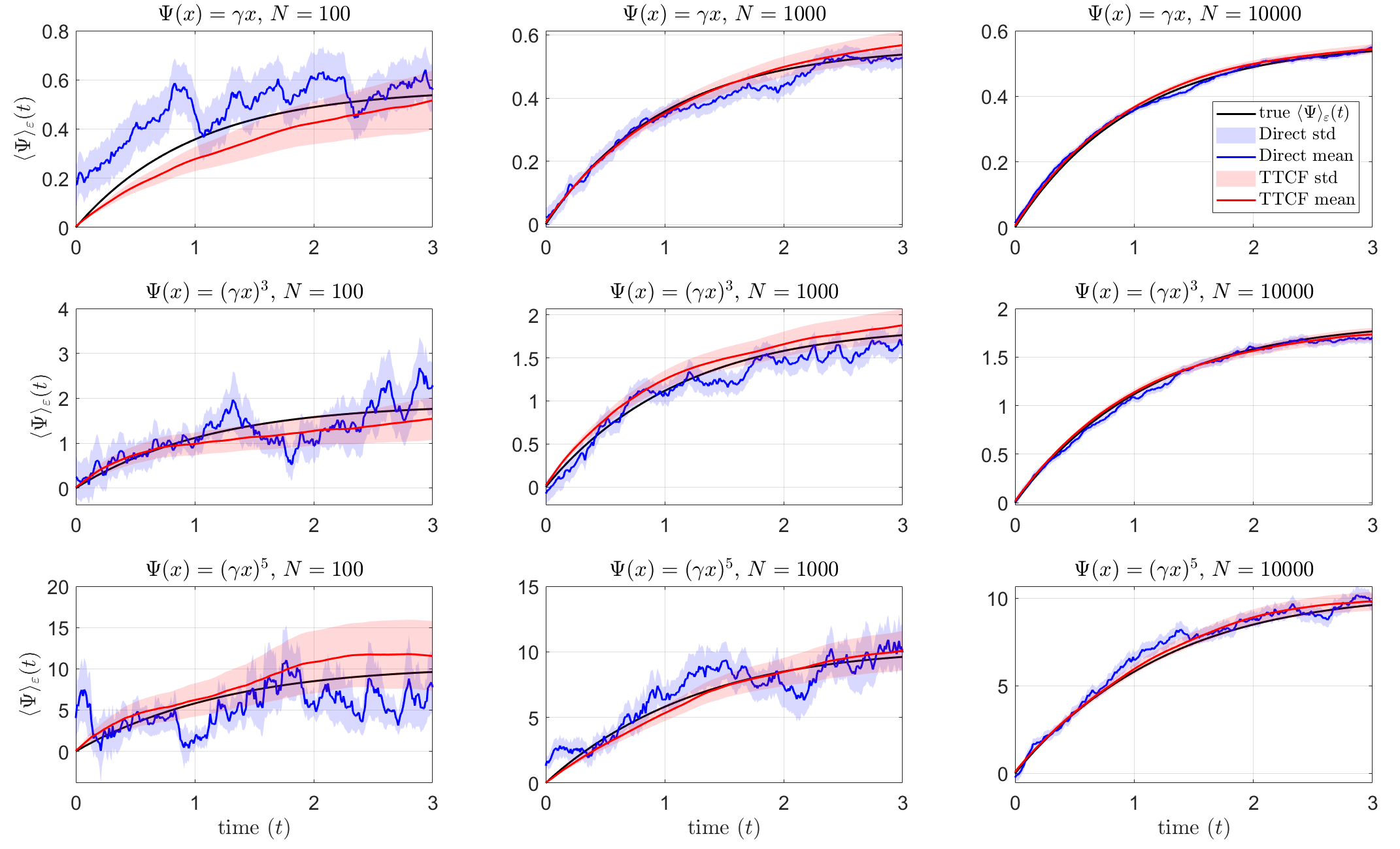}
	\caption{\label{fig:4}In blue we show the direct averages of the response of $\Psi_i$ relative to perturbations in $\varepsilon$ in Eq.~\eqref{eq:ou1d}. In red, we show the TTCF calculations. The observable analysed and number of ensemble members is indicated in the titles of each panel. The black lines are the ground truth computed using the Fokker-Planck equation associated with Eq.~\eqref{eq:ou1d}.}
\end{figure}

On Fig.~\ref{fig:4} we show the direct ensemble averages (in blue) and the TTCF method (in red) for a perturbation strength of $\varepsilon=0.1$ and noise intensity $\sigma = 0.25$. The ground truth shown in black is calculated by solving the Fokker-Planck equation \eqref{eq:fpe 2} relative to Eq.~\eqref{eq:ou1d} with an initial conditions given by $\rho_0$, and taking expectation values of $\Psi_i$, for $i=1,2,3$. The method is based on finite-differences \cite{CHANG19701}. Columns of the plot correspond to different numbers of ensemble members $N$ (indicated in the title) and rows correspond to the observables in question (indicated in the title). The shaded areas show the standard error at each time. As expected, direct averages converge to the true values as $N$ increases, although the rate at which this accuracy is improved is much lower compared to the TTCF method, where the approximations are smooth for all cases. The standard deviations, though are virtually uniform for the direct averages, whereas for the TTCF, as expected, they increase considerably as time increases. In both methods, though, as the number of ensemble members increases, the standard errors decrease, as expected.

\subsection{An OU process with rotational force}\label{sec:ou_2d}

Another level of complexity is added to Eq.~\eqref{eq:ou1d} by introducing non-conservative driving forces. For that it is necessary to add an extra dimension and consider the OU process in two-dimensions $\mathbf{x}(t)$:
\begin{equation}\label{eq:ou_2d}
    \mathrm{d}\mathbf{x}(t) = \mathbf{A}\mathbf{x}(t)\mathrm{d}t + \varepsilon \mathbf{f} +  \sigma \mathrm{d}\mathbf{W}_{t},
\end{equation}
where $\mathbf{A}$ is in $\mathbb{R}^{2\times 2}$, $\mathbf{f}$ is in $\mathbb{R}^2$, $\varepsilon>0$, $\sigma>0$ and $\mathbf{W}_t$ is a two-dimensional independent Wiener process. The matrix $\mathbf{A}$ is constructed as:
\begin{equation}
    \mathbf{A} = \begin{pmatrix} \hphantom{-}a & \hphantom{-}b \\ -b & \hphantom{-}a \end{pmatrix},
\end{equation}
which has eigenvalues $\lambda_{\pm} = a\pm i b$. If $a<0$ the unforced process is stable and has a Gaussian invariant measure given by:
\begin{equation}\label{eq:inv_mes_2d_ou}
    \rho_0(\mathbf{x}) =
\frac{-a}{\pi\sigma^2}
\exp\!\left( \frac{a}{\sigma^2}\,\|\mathbf{x}\|_2^2 \right),
\end{equation}
for $\mathbf{x}$ in $\mathbb{R}^2$. Notice that there is no presence of the parameter $b$ which is essentially introducing directionality and nonzero probability fluxes. The larger $b$, the stronger these fluxes are. Hence, we shall investigate the role of $b$ in computing ensemble averages. Since the invariant measure is known--- Eq.~\eqref{eq:inv_mes_2d_ou}---, we can readily compute $\Omega$:
\begin{equation}\label{eq:Omega_2d_ou}
    \Omega(\mathbf{x}) = -\frac{\nabla\cdot\left( \mathbf{f}\rho_0(\mathbf{x}) \right)} {\rho_0(\mathbf{x})} = -\frac{2a}{\sigma^2}\left( x_1 + x_2 \right),
\end{equation}
allowing us to apply the TTCF method. We investigate the following parameter configuration: $a=-1$, $\sigma=0.4$, $\mathbf{f}=(1,1)^{\ast}$ and $\varepsilon = 0.1$.

\begin{figure}[H]
\centering
\includegraphics[scale=0.44]{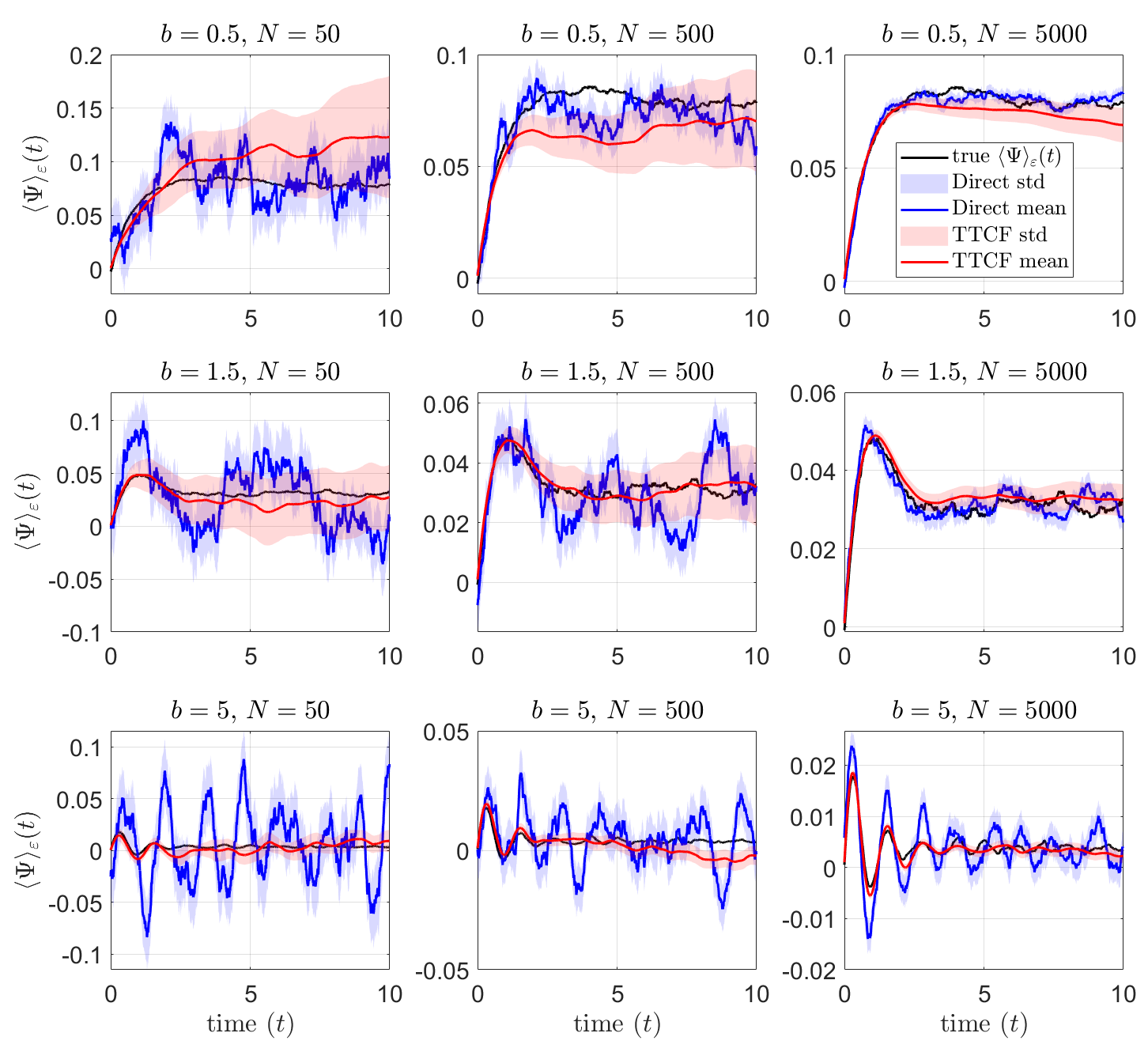}
\caption{\label{fig:2}Each panel shows the same comparison of direct averages and TTCF as Fig.~\eqref{fig:4}, but this time for Eq.~\eqref{eq:ou_2d} and different values of both $N$ and $b$. Here, we show the response of the observable $\Psi(\mathbf{x})=x_1$, in the system Eq.~\eqref{eq:ou_2d} with respect to the forcing $\varepsilon\mathbf{f}$. Each row corresponds, respectively, to $b=0.5,1.5$ and $5$ and each column corresponds to $N=50,500$ and $5000$. The parameters are selected to be: $a=-1$, $\sigma=0.4$, $\mathbf{f}=(1,1)^{\ast}$ and $\varepsilon = 0.1$}
\end{figure}

On Fig.~\eqref{fig:2} we show three-by-three panels with rows corresponding to three different choices of $b=0.5,1.5,5$, and columns corresponding to $N=50,500,5000$ ensemble members in the approximation of direct averages (in blue) and TTCF runs (in red). The black curves show the ground truth corresponding to direct averages with $N=50,000$. For $b=0.5$, while the SNR of the TTCF method is higher than that of the direct averages, it does not necessarily provide a more reliable estimation unless $N$ is higher. However, as $b$ increases we observe two aspects. First, as shown in the second row, the transient growth due to non-equilibrium fluxes is very well captured by the TTCF method and for large times, it appears that direct averages still fail to accurately capture the ground truth value. Secondly, for strong non-conservative forces $b=5$, direct averages dramatically fail to capture the time dependent averages, yielding a significant out-performance of the TTCF methodology. 

\subsection{The Lorenz 96 system}
The single-scale Lorenz 96 (L96) model is a widely used idealized system for studying
nonlinear dynamics, chaos, predictability, and data assimilation in geophysical flows.
It consists of $L$ prognostic variables $x_i$, for $i=1,\ldots,L$, each representing a scalar
quantity arranged on a periodic one-dimensional lattice.
In spite of its simplicity, the model exhibits key features of atmospheric dynamics such as
advection-like nonlinearities, forcing, and dissipation. To make the system a bit more realistic, we shall consider that the energy input of the system is a stochastic forcing, mimicking the small fluctuations in the solar irradiance.
The governing equations of the single-scale stochastic L96 model are
\begin{equation}\label{eq:L96}
\dot{x}_i
= (x_{i+1} - x_{i-2})\,x_{i-1} - x_i + F +\sigma \dot{W}_t,
\end{equation}
where $i=1,\ldots,L$ indices are taken modulo $L$ to satisfy the periodicity of the domain. The quadratic term represents the advection, the linear term $-x_i$ is the dissipation, and the constant forcing $F$ injects energy into the system. Here we take $L=20$ and $F=8$, which leads to a well reported chaotic regime in case of $\sigma=0$. The constant $\sigma>0$ represents the fluctuation intensity of the energy input. Small and large values of $\sigma$ represent, respectively, weak and strong fluctuation regimes in the model.

The response of this model has been investigated under many points of view. Linear response and Green's functions were obtained from impulse-response experiments to determine its dynamical susceptibilities \cite{lucarini2011statistical,Tomasini2021}, fluctuation-dissipation relations \cite{majda_abramov_book} and response to stochastic forcing \cite{Lucarini2012}. Here, instead, we aim at computing the ensemble averages to the response to perturbations of the parameter $F\mapsto F+\varepsilon$--- for $\varepsilon>0$--- and compare the estimation of the response using direct averages versus the TTCF method.

Unlike the OU process, the L96 model, just as many other dynamical systems, does not have an explicit analytic expression for the invariant measure $\rho_0$. Yet, we know that since we are dealing with an elliptic diffusion process and suitable conditions are verified for the drift term, then the invariant measure is absolutely continuous with respect to Lebesgue and the density obeys Eq.~\eqref{eq:fpe 2} \cite{pavliotisbook2014}. The absence of an explicit form for $\rho_0$ is a crucial barrier in the direct application of fluctuation-dissipation prediction methods, since they require the construction of the dissipation function $\Omega(\cdot)$, analogously  to that done in Eqs.~\eqref{eq:Omega_1d_ou} and \eqref{eq:Omega_2d_ou}. In the sections below we adopt  approximation methodologies for estimating $\rho_0$ in order to perform TTCF computations. We focus on a simple Gaussian approximation and on kernelized approximations inspired by extended dynamical mode decomposition algorithms \cite{Zagli2025}.

\subsubsection{Gaussian approximation}\label{sec:gaussian1}

A close look to time-series produced by Eq.~\eqref{eq:L96} reveals nontrivial statistical features product of an underlying chaotic attractor, yet the overall view shows uni-modal and normal behavior. Hence, a natural way for approximating the function $\Omega(\cdot)$ is by assuming that the invariant measure of the system $\rho_0$ can be represented as a Gaussian distribution:
\begin{equation}\label{eq:pm_l96}
	\rho_0(\mathbf{x}) =
	\frac{1}{\sqrt{(2\pi)^L \det(\Sigma)}}\,
	\exp\!\left( -\frac{1}{2}(\mathbf{x}-\boldsymbol{\mu})^\ast \Sigma^{-1} (\mathbf{x}-\boldsymbol{\mu}) \right),
\end{equation}
where $\boldsymbol{\mu}$ and $\Sigma$ are the mean vector and covariance matrix estimated from long timeseries of the unforced model. Under this setting, the function $\Omega(\cdot)$ becomes a linear function of the state variables $x_i$:
\begin{equation}\label{eq:L96_dissipation_function}
	\Omega(\mathbf{x}) = -  \mathbf{1}_L^{\ast}\Sigma^{-1}\left( \mathbf{x} - \boldsymbol{\mu} \right),
\end{equation}
where $\mathbf{1}_L$ is the $L$-dimensional vector full of ones. Note the similarity to what was obtained in previous sections for the OU processes.

\subsubsection{Kernelized approximation}\label{sec:kernel_rho_0}

The Gaussian approximation is the simplest ansatz for the invariant distribution of a system. However, the slightest deviations from pure Gaussianity might compromise the estimation of the response via TTCF. Here we propose an kernelized extension of the Gaussian distribution for the estimation of the dissipation function $\Omega(\cdot)$. 

We aim at approximating the dissipation function as a combination of kernel functions $\kappa(\cdot, \cdot)$ evaluated at $M$ time series data points $\{\mathbf{x}\}_{i=1}^M$ with statistics obeying the invariant measure $\rho_0$. In other words we seek to minimize the following quantity:
\begin{equation}
    \left| \Omega(\mathbf{x}) - \sum_{i=1}^M\xi_i \kappa(\mathbf{x},\mathbf{x}_i) \right|,
\end{equation}
for points $\mathbf{x}$ distributed according to $\rho_0$. The vector of coefficients $\boldsymbol{\xi} = (\xi_i) $ are thus obtained in the following form:
\begin{equation}\label{eq:xi_formula}
    \boldsymbol{\xi} = \mathbf{H^{\dagger}\boldsymbol{\Delta}}, 
\end{equation}
where the superscript ``$\dagger$" denotes the Moore-Penrose inverse, and we have defined 
\begin{equation} \label{H_definition}
    H_{ij} = \int\kappa(\mathbf{x},\mathbf{x}_i)\kappa(\mathbf{x},\mathbf{x}_j)\rho_0(\mathbf{x})\mathrm{d}\mathbf{x}\approx \frac{1}{M}\sum_{k=1}^M\kappa(\mathbf{x}_k,\mathbf{x}_i)\kappa(\mathbf{x}_k,\mathbf{x}_j),
\end{equation}
and $\boldsymbol{\Delta}= (\Delta_i)$ -- a vector whose $i$-th component obeys the following chain of relations:
\begin{subequations}\label{eq:delta_derivation}
\begin{align}
    \Delta_i &= \int \kappa(\mathbf{x},\mathbf{x}_i) \Omega (\mathbf{x})\rho_0(\mathbf{x})\mathrm{d}\mathbf{x}= \int G(\mathbf{x})\cdot \nabla_1 \kappa(\mathbf{x},\mathbf{x}_i) \rho_0(\mathbf{x})\mathrm{d}\mathbf{x} \\ &\approx \frac{1}{M}\sum_{k=1}^MG(\mathbf{x}_k)\cdot \nabla_1 \kappa(\mathbf{x}_k,\mathbf{x}_i)  = \frac{1}{M}\sum_{k=1}^M\mathbf{1}_L\cdot \nabla_1\kappa(\mathbf{x}_k,\mathbf{x}_i), 
\end{align}
\end{subequations}
where the subscript ``1" indicates the first argument of $\kappa(\cdot,\cdot)$. In the approximation we have used the ergodicity assumption and the fact that $G(\cdot)$ is a constant vector in the perturbation problem considered in Eq.~\eqref{eq:L96} by means of $F\mapsto F+\varepsilon$. 
Note that \eqref{eq:delta_derivation} means that the coefficients $\mathbf{\xi}_i$ may be computed without knowledge of the form of $\rho_0(\mathbf{x})$, if we instead compute $\kappa(\mathbf{x}_i,\mathbf{x}_j)$ and its gradient at every point on a long trajectory. 

While formulas Eq.~\eqref{H_definition} and \eqref{eq:delta_derivation} are general, in the present study we use Gaussian radial basis functions (GRBFs) as kernel functions with bandwidth $\eta>0$. This amounts to saying that the structure of the dissipation function will be linear in $\mathbf{x}$. Indeed,
\begin{equation}
    \nabla_1 \kappa(\mathbf{x},\mathbf{x}_i) = \frac{1}{2\eta^2}(\mathbf{x}-\mathbf{x}_i)\kappa(\mathbf{x},\mathbf{x}_i),
\end{equation}
for every $i=1,\ldots,M$. Using this, we can construct $\Omega(\cdot)$ through the linear combination of kernel functions 
\begin{equation}
    \Omega(\mathbf{x}) = \sum_{i=1}^M\xi_i \kappa(\mathbf{x},\mathbf{x}_i).
\end{equation}


\subsubsection{Computing the response of observables}

For the chaotic parameter configuration (with $L=20$) and the specified constant forcing $\varepsilon$ in Eq.~\eqref{eq:L96}, we target computing the response of the observable function $\Psi_j(\mathbf{x})$ defined as:
\begin{equation}\label{eq:observable_l96}
	\Psi_j(\mathbf{x}) = \frac{1}{jL}\sum_{i = 1}^Lx^{j}_i - \frac{1}{jL} \int \sum_{i = 1}^Lx^{j}_i \rho_0(\mathbf{x})\mathrm{d}\mathbf{x},
\end{equation}
for $j=1,2,3$. Note that the integral term in the definition of $\Psi_j$ implies that $\langle \Psi_j\rangle_0 = 0$, for $j=1,2,3$.

The first step is to sample the actual invariant measure of the system when no forcing is applied: $\varepsilon=0$. For that we integrate Eq.~\eqref{eq:L96} for $5\cdot 10^7$ time units with a time step of $ 10^{-2}$ time units, employing an Euler-Maruyama update. A transient of $50$ time units was removed prior to recording the time-series to ensure convergence in the steady state. From such a time series the sample covariance matrix $\Sigma$ and sample mean $\boldsymbol{\mu}$ are computed to obtain the Gaussian approximation of $\rho_0$ according to Eq.~\eqref{eq:pm_l96}. Then, $N>0$ equidistant-in-time snapshots are saved, so that they constitute the initial sampling of the invariant state $\rho_0$ shown in Eq.~\eqref{eq:pm_l96}. Such sampling is independent since the time-separation between snapshots considered in the experiments--- specified below--- exceeds five time units, which is far beyond the characteristic decorrelation timescales of the model. Note that if one assumes an exact Gaussian approximation via Eq.~\eqref{eq:pm_l96}, one could also directly and independently sample such a distribution. Regarding the GBRF kernel basis estimation shown in Eq.~\eqref{sec:kernel_rho_0}, we employed a time series consisting of $2\cdot 10^3$ snapshots with a time step of $0.1$ time units. A bandwidth of $\eta = 20$ is used. As a next step, formula \eqref{eq:xi_formula} was employed to find the expansion of $\Omega(\cdot)$ in terms of the kernel basis. That allows us to directly evaluate the function $\Omega(\cdot)$ and compute the TTCF estimator $C_t$ shown in Eq.~\eqref{eq:estimators_R_t_C_t}.

\begin{figure}[H]
	\centering
	\includegraphics[scale=0.26]{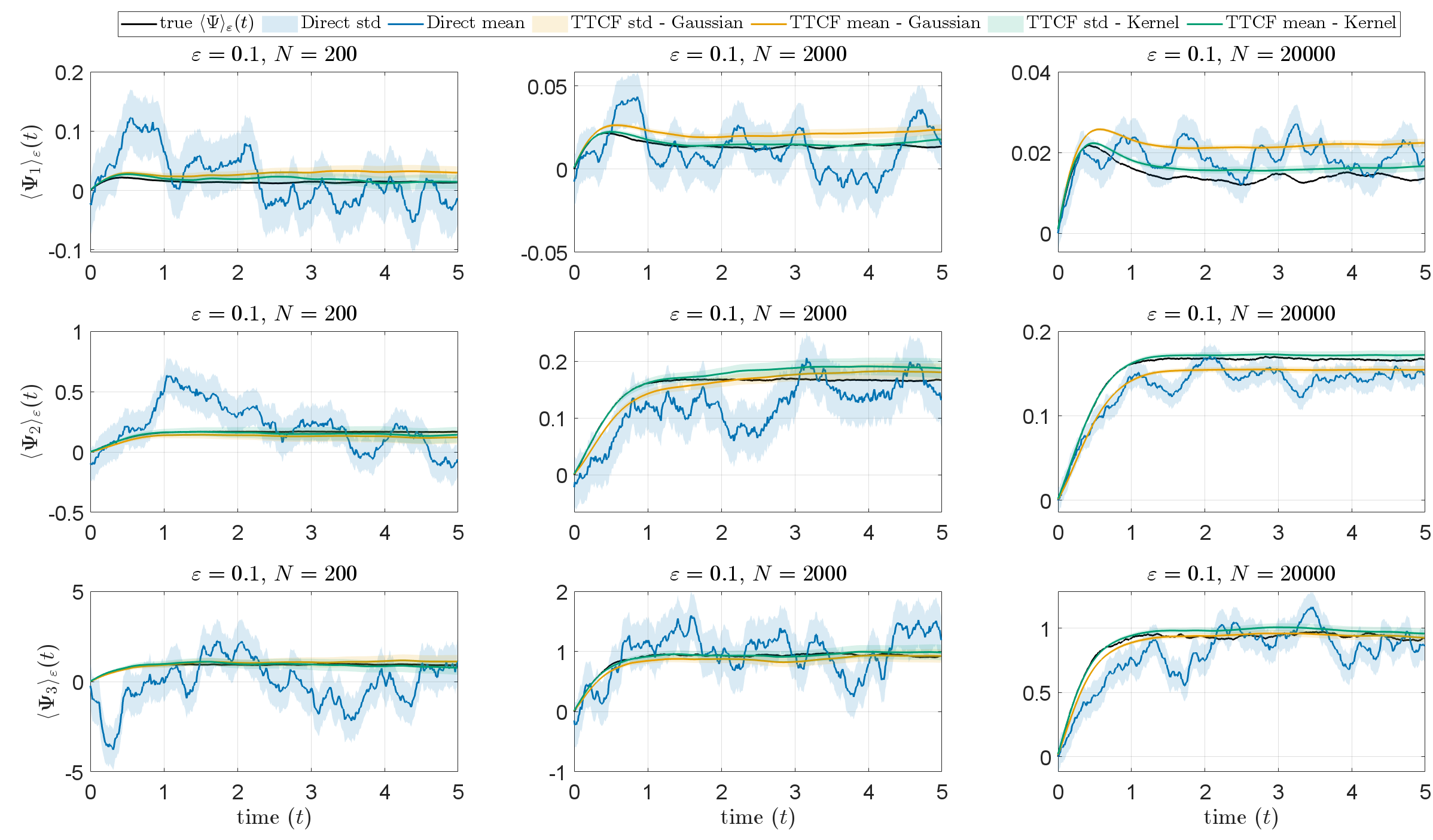}
	\caption{\label{fig:L96_pm_sigma_25_eps_005}\textbf{Direct averages vs TTCF: $\varepsilon=0.1$.} We show the response of the observable $\Psi_j$ defined Eq.~\eqref{eq:observable_l96} in response to $F\mapsto F +\varepsilon$. In blue, we show the direct averages, in yellow, the TTCF estimation using a Gaussian approximation and in green, the TTCF estimation employing the KDMD method. The number of ensemble members is indicated in the titles together with the forcing strength $\varepsilon$.  The black lines correspond to the ground-truth computed from direct averages using $10^7$ independent ensemble members.}
\end{figure}

\begin{figure}[H]
	\centering
	\includegraphics[scale=0.26]{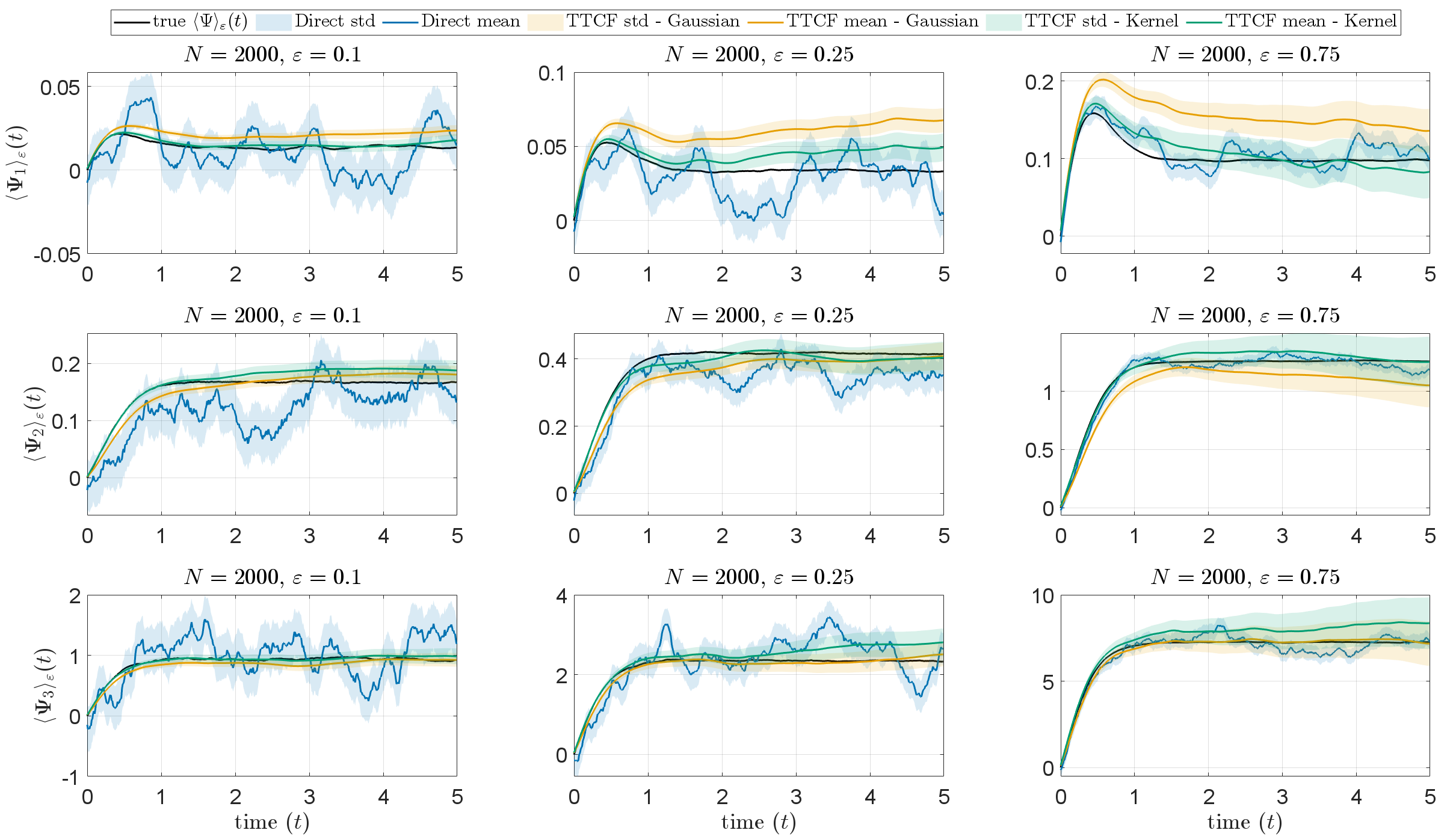}
	\caption{\label{fig:L96_fixedN}\textbf{Direct averages vs TTCF: $\varepsilon=0.1$.} We show the response of the observable $\Psi_j$ defined Eq.~\eqref{eq:observable_l96} that results form $F\mapsto F +\varepsilon$. In blue, we show the direct averages, in yellow, the TTCF estimation using a Gaussian approximation and in green, the TTCF estimation employing the Kernel method together with the forcing strength $\varepsilon$. The number of ensemble members is indicated in the titles.  The black lines correspond to the ground-truth computed from direct averages using $10^7$ independent ensemble members.}
\end{figure}

Figure~\ref{fig:L96_pm_sigma_25_eps_005} shows (in blue) the computation of the direct averages of the response of the system to $F\mapsto F +\varepsilon$ of the observable $\Psi_j$ defined in Eq.~\eqref{eq:observable_l96}. The yellow and green lines are computed using the TTCF approach, both for the Gaussian and kernelized methods. The observable under scrutiny is indicated in the $y$-axis. The black line indicates our ground truth computed with direct averages using $10^7$ snapshots separated in time by five time units. The shaded  areas around their respective curves show the standard error computed from the $N$ ensemble members. The same plotting scheme was used in Fig.~\ref{fig:L96_fixedN}, where instead the number of samples $N$ was fixed to $2000$. Instead, the forcing value $\varepsilon$ is increased successively ($\varepsilon=0.1,0.25,0.75$) to illustrate the role of the strong-forcing regime.

The dependence on the number of samples $N$ illustrates one of the main advantage of the TTCF method. Throughout the columns of Fig.~\ref{fig:L96_pm_sigma_25_eps_005}, we observe that the TTCF computation (both Gaussian and kernelized approach) provides smoother curves, product of the improved SNRs. Even for the case of low $N$ the TTCF is able to provide a better estimate of the response, especially at early times. This method mitigates spontaneous deviations that arise from a poor sampling of the initial stationary measure that are seen in the direct averages (blue curves), particularly for $N=200$ and $2000$. As $N$ increases, the standard error of the direct averages and TTCF, naturally, decreases. However, while the standard error of the direct averages remains practically constant throughout the integration window, the TTCF grows with time, although at initial times the TTCF mitigates the totality of the fluctuations, as seen in previous sections. One must note, nevertheless, that in the limit of small $N$, direct averages will suffer more--- vanishing SNR---, particularly, at early times, where TTCF methods clearly perform better, consequence of having a non-zero SNR.

Both Gaussian and kernelized approached yielded smooth responses, although the main improvement of the kernelized approach with respect to its Gaussian counterpart is in resolving the response of $\Psi_1$ at early times. Indeed, three observables where investigated in these numerical experiments--- see Eq.~\eqref{eq:observable_l96}--- for which the computation of their respective responses differed. While the response of $\Psi_2$ and $\Psi_3$ where monotonous in time, $\Psi_1$ displayed a clear transient feature, this is, a relatively sharp increase in its expectation value before a relaxation to its perturbed steady-state value. This transient feature was especially hard to capture using the Gaussian approach, whereas the kernelized method did capture the transient response more accurately. 

It was briefly mentioned in the text above that stronger forcing can lead to better capturing the signal of the perturbation, as was also demonstrated in \cite{MAFFIOLI2024109205}. This is clearly seen in the present experiments. Figure~\ref{fig:L96_fixedN} shows the results for increasing values of $\varepsilon$, but keeping the number of samples fixed (indicated in the titles). It is when $\varepsilon=0.75$ (strongest-forcing regime) that direct averages start to converge to the actual response and outperform the TTCF methods. Consequently, the TTCF algorithm exhibits an advantage with respect to taking direct averages in the weak forcing regime, as analytically shows in Eqs.~\eqref{eq:da_limit_epsilon} and \eqref{eq:ttcf_limit_epsilon}.

\section{Conclusions}\label{conclusions}
In this work we have examined the applicability of the transient time correlation function (TTCF) method \cite{morris_1987,evans_morris_1990} in the context of chaotic and stochastic systems out of equilibrium. Original formulations of TTCF come from the study of  thermostatted molecular systems for which invariant measures and dissipation functions are written in terms of the mechanical energy of the system \cite{evans_morris_1990}. This implies three things: 1) linear response algorithms via fluctuation-dissipation theorems are available from sampling the unforced system \cite{kubo}, 2), the TTCF method is exact and readily applicable to provide better signal-to-noise ratios (SNRs) when computing ensemble averages \cite{morris_1987,MAFFIOLI2024109205} and 3), it allows to formulate closed equations for the exact response to all orders of the perturbation strength \cite{colangeli_2025}.


We have hereby derived the TTCF for  nonequilibrium systems: formulas are obtained both for a general class of diffusion processes and for general Markov chains (see Appendix~\ref{TTCFMarkov}). A spectral representation of the evolution operators allowed for the investigation of the performance of the TTCF in the temporal dimension. We derived an explicit spectral expression for the signal-to-noise ratio (SNR) at early and long times and, furthermore, the dependence on the choice of observable was highlighted. In particular, we show how choosing observables projecting in fast Koopman eigenfunction yield worse response estimates resulting from poor SNRs.


For nonequilibrium systems featuring strong departure from gradient dynamics, we encountered a remarkable performance of the TTCF method as compared to direct averages. We examined a 2D Ornstein-Uhlenbeck process that includes a rotational component. We showed that increasing the strength of the rotational force of the unperturbed system led to oscillations that could be hardly if at all captured by  simply taking direct averages. As opposed to that, the TTCF was much more effective in damping the oscillations and converging to the ground truth. Both analytical--- see App.~\ref{rotation}--- and numerical results are restricted such a linear process and extending the claims to general systems requires more work. {\color{black} While the spectral expansions apply to any nonlinear system with a quasi-compact evolution operator \cite{chekroun2019c}, the analytical derivation of the response dependence on non-conservative forces was done here exclusively for the OU process which is linear and yields linear responses--- see App.~\ref{rotation}---. Due to the linearity of the base system, the signal to noise ratio will depend linearly on the perturbation strength. This, however, cannot be directly extrapolated to higher order nonlinear systems.}

We applied the TTCF to the stochastic L96 system. The invariant measure of the reference system is not explicitly known and the deterministic dynamics is forced, and dissipative, and chaotic, placing ourselves out of thermodynamic equilibrium. The main barrier for the implementation of TTCF stems from the need of an analytical expression for the dissipation function $\Omega(\cdot)$--- see definition above Eq.~\eqref{eq:derivative_epsilon}---. This requires an analytical treatment of the invariant measure and its derivatives. Works in the literature have fitted data from the reference state to a wide catalog of distributions. These include, most importantly, Gaussian mixtures \cite{majda_abramov_book}, which yield analytic expressions for $\Omega(\cdot)$. Here, the TTCF method was implemented with a single Gaussian approximation of the stationary state and a kernelized method based on radial basis functions, out of which $\Omega(\cdot)$ can be directly computed. In both cases, the corresponding TTCF evaluation yielded smooth response functions for any number of initial samples. This raises the question of the applicability of the present results to larger, high-dimensional systems. An immediate question would be to assess model parameter sensitivity, which is can be recast as a response to small parameter changes.

Concluding, we make a strong case that for a sparsely sampled stationary states, the TTCF method provides a clearer response signal to forcings compared to direct averages.  
Along these lines, a time-dependent version of TTCF was formulated recently in \cite{Iannella2023-tb}, yet a detailed numerical investigation remains to be done. Next stages should focus in developing a time-dependent formulation using the Fokker-Planck operator approach and implementing it to high-dimensional estimation of the response to time-modulated forcings, particularly, applications in high-dimensional geophysical models should be explored.

\paragraph{Acknowledgements}\quad MSG would like to thank Luca Maffioli for dedicating time to explain the applications of TTCF methods in molecular physics. VL and MSG acknowledge the partial support provided by the Horizon Europe Projects Past2Future (Grant No. 101184070). VL acknowledges the partial support provided by the Horizon Europe project ClimTIP (Grant No. 100018693), by the ARIA SCOP-PR01-P003—Advancing Tipping Point Early Warning AdvanTip project. NZ and VL are also supported by the European Space Agency Project PREDICT (Contract 4000146344/24/I-LR). 

\pagebreak
\appendix

\section{Derivation of $\mathrm{Var}[C_t]$}\label{app:derivation_variance}

Here we show the derivation of Eq.~\eqref{eq:snr_ttcf}. Departing from Eq.~\eqref{eq:mean_c_t}, where we computed the first moment, we compute the second uncentered moment of $C_t$:

\begin{subequations}
    \begin{align}
        \mathbb{E}\left[ C_t^2 \right] &= \frac{\varepsilon^2}{N^2}  \sum_{i,j=1}^N \int_{0}^t\int_{0}^t\mathbb{E}\left[\Psi(X_s^{(i)})\Omega(X_i)\Psi(X_u^{(j)})\Omega(X_j) \right]\mathrm{d}s\mathrm{d}u. \\&= \frac{\varepsilon^2}{N^2}  \sum_{i=1}^N \int_{0}^t\int_{0}^t\mathbb{E}\left[(P^{\varepsilon}_s\Psi)(X_i)\Omega(X_i)(P^{\varepsilon}_u\Psi)(X_i)\Omega(X_i) \right]\mathrm{d}s\mathrm{d}u \\ &+ \frac{\varepsilon^2}{N^2}\sum_{i\neq j =1}^N\int_{0}^t\int_{0}^t\mathbb{E}\left[(P^{\varepsilon}_s\Psi)(X_i)\Omega(X_i)\right]\mathbb{E}\left[(P^{\varepsilon}_u\Psi)(X_j)\Omega(X_j)\right]\mathrm{d}s\mathrm{d}u\\        &= \frac{\varepsilon^2}{N}  \int_{0}^t\int_{0}^t\mathbb{E}\left[(P^{\varepsilon}_s\Psi)(x)\Omega(x)(P^{\varepsilon}_u\Psi)(x)\Omega(x) \right]\mathrm{d}s\mathrm{d}u + \frac{\varepsilon^2(N-1)}{N}\int_{0}^t\int_{0}^t\Gamma^\varepsilon(s)\Gamma^\varepsilon(u)\mathrm{d}s\mathrm{d}u,
    \end{align}
\end{subequations}
where we have used the tower rule in the second equality, the independence of the variables $\{X_i\}_{i=1}^N$ and we have introduced 
$\Gamma^\varepsilon (s) =\mathbb{E}\left[(P^{\varepsilon}_s\Psi)(X)\Omega(X)\right]$.
Since the square of the first moment of the TTCF estimator is
\begin{equation}
    \mathbb{E}\left[ C_t \right]^2 = \varepsilon^2\left(\int_{0}^t \Gamma^\varepsilon (s)\mathrm{d}s\right)\left(\int_{0}^t \Gamma^\varepsilon (u)\mathrm{d}u\right) = \varepsilon^2\int_{0}^t\int_0^t\Gamma^\varepsilon(s)\Gamma^\varepsilon(u)\mathrm{d}s\mathrm{d}u,
\end{equation}
we can evaluate its variance
\begin{equation}\label{eq:var_c}
    \mathrm{Var}\left[ C_t \right] = \frac{\varepsilon^2}{N}\int_{0}^t\int_0^t \bigg(\mathbb{E}\left[(P^{\varepsilon}_s\Psi)(x)\Omega(x)(P^{\varepsilon}_u\Psi)(x)\Omega(x) \right] - \Gamma(s)\Gamma(u) \bigg)\mathrm{d}s\mathrm{d}u,
\end{equation}
which corresponds to a lagged autocovariance of the stochastic process $\Psi(X_t)\Omega(X)$ in the perturbed flow \eqref{eq:sto ode 2}. Finally, the SNR of $C_t$ is
\begin{equation}
    \mathrm{SNR}[C_t] = \sqrt{N}\frac{|\int_{0}^t \Gamma^\varepsilon (s)\mathrm{d}s|}{\sqrt{\int_{0}^t\int_0^t \bigg(\mathbb{E}\left[(P^{\varepsilon}_s\Psi)(x)\Omega(x)(P^{\varepsilon}_u\Psi)(x)\Omega(x) \right] - \Gamma^\varepsilon(s)\Gamma^\varepsilon(u) \bigg)\mathrm{d}s\mathrm{d}u}}.
\end{equation}

\section{The Ornstein-Uhlenbeck process: the effects of rotation}\label{rotation}

In this appendix we want to derive exact relations to determine the role of a strong rotational component and large constant forcing in an $N$-dimensional OU process. We depart from the following stochastic differential equation:
\begin{equation}\label{eq:OU_app}
    \dot{\mathbf{x}}(t) = \mathbf{A}\mathbf{x}(t) + \delta \mathbf{B}\mathbf{x}(t) + \varepsilon \mathbf{f} + \sigma \dot{\mathbf{W}}_t,
\end{equation}
where $\mathbf{A}=(A_{ii})$ is diagonal in $\mathbb{R}^{N\times N}$ with negative entries, $N$ is even, $\varepsilon,\sigma>0$, $\mathbf{f}$ is in $\mathbb{R}^{N}$, $\mathbf{W}_t$ is an $N$-dimensional independent Wiener process and $\mathbf{B}$ us a block diagonal matrix in $\mathbb{R}^{N\times N}$ where each block $\mathbf{B}_j$ is defined as:
\begin{equation}
	\mathbf{B}_j = \begin{bmatrix}
		\hphantom{-}0 & \hphantom{-}\omega_j \\ -\omega_j & \hphantom{-}0,
	\end{bmatrix}
\end{equation}
with $\omega_j>0$, where $j=1,\ldots,N/2$. It follows then that the eigenvalues of the matrix $\mathbf{A} + \delta \mathbf{B}$ are $\{A_{ii}\pm \omega_i\}_{i=1}^{N}$.

We want to study the response to the forcing $\varepsilon \mathbf{f}$ of a linear observable $\Psi$ of the form:
\begin{equation}
	\Psi(\mathbf{x}) = \sum_{i=1}^N \alpha_i x_i,
\end{equation}
where $\boldsymbol{\alpha} = (\alpha_i)$ is a real $N$-dimensional vector. We then have:
\begin{equation}
\mathbb{E}\left[ \Psi\right](t) = \varepsilon \int_0^t e^{\left( \mathbf{A} + \delta \mathbf{B} \right)s}\mathbf{f}\mathrm{d}s.
\end{equation}
and the covariance matrix $K(t)$ reads as:
\begin{subequations}    
	\begin{align}
	K(t)&=e^{\left( \mathbf{A} + \delta \mathbf{B} \right)t}K(0)e^{\left( \mathbf{A} + \delta \mathbf{B} \right)t} + \sigma^2 \int_{0}^te^{\left( \mathbf{A} + \delta \mathbf{B} \right)s}e^{\left( \mathbf{A} + \delta \mathbf{B} \right)^{\ast}s}\mathrm{d}s \\ &= e^{\left( \mathbf{A} + \delta \mathbf{B} \right)t}K(0)e^{\left( \mathbf{A} + \delta \mathbf{B} \right)t} + \sigma^2 \int_{0}^t e^{2\mathbf{A}s}\mathrm{d}s,
	\end{align}
\end{subequations}
where we have used the fact that $\mathbf{A}$ and $\mathbf{B}$ commute and that $\mathbf{B} + \mathbf{B}^{\ast}\equiv 0$. Hence, the variance of $\Psi$ is independent of $\varepsilon$. Consequently, the SNR of computing the direct averages of $\Psi$ is:
\begin{equation}
	\mathrm{SNR}(D(t)) = \frac{\varepsilon \int_0^t e^{\left( \mathbf{A} + \delta \mathbf{B} \right)s}\mathbf{f}\mathrm{d}s}{\boldsymbol{\alpha}^{\ast}K(t)\boldsymbol{\alpha}},
\end{equation}
which increases proportionally to the perturbation strength $\varepsilon$. This shows that in the limit of strong forcing, direct averages are enough to obtain a good signal.

Now, we shall derive the role of the rotation strength $\delta$. For that we write a cleaner expression for the numerator of $\mathrm{SNR}(D(t))$. Since $\mathbf{A}$ and $\delta \mathbf{B}$ commute, we have
\begin{equation}
e^{(\mathbf{A}+\delta \mathbf{B})t} = e^{\mathbf{A} t} e^{\delta \mathbf{B} t}.
\end{equation}
The exponential of each block $\mathbf{B}_j s$ is a rotation matrix:
\begin{equation}
e^{\delta\mathbf{B}_j s} = \begin{bmatrix} \hphantom{-}\cos(\delta\omega_j s) & \hphantom{-}\sin(\delta\omega_j s) \\ -\sin(\delta\omega_j s) & \hphantom{-}\cos(\delta\omega_j s) \end{bmatrix}.
\end{equation}
Then, 
\begin{align}\label{eq:formula_mean_OU}
\int_0^t \boldsymbol{\alpha}^\ast e^{(\mathbf{A}+\delta \mathbf{B})s} \mathbf{f} \mathrm{d}s
		&= \sum_{j=1}^{N/2} \int_0^t e^{A_{2j-1,2j-1} s} \Big[ 
		\alpha_{2j-1} \big(f_{2j-1} \cos(\delta\omega_j s) + f_{2j} \sin(\delta\omega_j s)\big) \nonumber\\
		&\quad \quad \quad + \alpha_{2j} \big(- f_{2j-1} \sin(\delta\omega_j s) + f_{2j} \cos(\delta\omega_j s)\big)
		\Big] \, \mathrm{d}s,
	\end{align}
where $\boldsymbol{\alpha}=(\alpha_i)$. Now, we have to integrate multiple exponentials times sinusoidal functions. In general, we have the following standard integrals to solve recursively:
\begin{subequations}
\begin{align}\label{eq:integrals_sin_exp}
	&\int_{0}^{t} e^{\lambda s}\sin(\omega s) \mathrm{d}s = e^{\lambda t}\frac{\lambda \sin(\omega t)-\omega \cos(\omega t)}{\lambda^2 + \omega^2} - \frac{\omega}{\lambda^2 + \omega^2} \\
	&\int_{0}^{t} e^{\lambda s}\cos(\omega s) \mathrm{d}s = e^{\lambda t}\frac{\lambda \sin(\omega t)+\omega \cos(\omega t)}{\lambda^2 + \omega^2} - \frac{\lambda}{\lambda^2 + \omega^2} .
\end{align}
\end{subequations}
Hence, a factor of $(A_{jj}+(\delta\omega_j)^2)^{-1}$ appears in the computation of the mean of $\Psi$. Thus, in the presence of a strong rotational force, namely, for $\delta \gg 1$, the we have that the $\mathrm{SNR}(D(t))$ is written as:
\begin{equation}\label{eq:snr_D_delta}
	\mathrm{SNR}(D(t)) = \left( \boldsymbol{\alpha}^{\ast}K(t)\boldsymbol{\alpha}\right)^{-1}\sum_{j=1}^N\frac{\kappa_j(t)}{\delta^2 \left(\left(\frac{\lambda_j}{\delta}\right)^2 + \omega_j^2\right)}\sim \mathcal{O}(\delta^{-2}),
\end{equation} 
where $\kappa_j(\cdot)$ is a time dependent function such that $\lim _{\delta \rightarrow \infty}\kappa_j(\cdot)/\delta^2 = 0$. Consequently, the signal to noise ratio of computing the average of a linear functional of the OU process in the presence of constant forcing decreases with the strengthening of the rotational force.

It remains to compute the scaling of the TTCF method with respect to $\delta$. Since the expression is more convoluted, we depart from the weak forcing approximation, namely, Eq.~\eqref{eq:ttcf_limit_epsilon}. In this case, the numerator for $D(t)$ and $C(t)$ is equal to Eq.~\eqref{eq:formula_mean_OU}. Then,
\begin{equation}\label{eq:numerator_derivation}
    \int_{0}^t\Gamma(s)\mathrm{d}s = \int_0^t \boldsymbol{\alpha}^\ast e^{(\mathbf{A}+\delta \mathbf{B})s} \mathbf{f}=\sum_{j=1}^N\frac{\kappa_j(t)}{\delta^2 \left(\left(\frac{\lambda_j}{\delta}\right)^2 + \omega_j^2\right)} \mathrm{d}s.
\end{equation}

We now have to examine what happens with the denominator in Eq.~\eqref{eq:ttcf_limit_epsilon}, which involves the correlation function $\chi_0(\cdot)$. For that we project the function $\chi_0(\cdot)$ as a sum of the first $M$ Koopman eigenpairs:
\begin{equation}
\chi_0(t) \approx \sum_{\ell=1}^{M} 
e^{\mathrm{Re}(\lambda_\ell) |t|} \Big[ \gamma_\ell \cos( \delta \mathrm{Im}(\lambda_\ell) t) + \xi_\ell \sin( \delta \mathrm{Im}(\lambda_\ell) t) \Big],
\end{equation}
where the complex value $\lambda_\ell$ is the $\ell$-th eigenvalue of the Koopman operator associated with Eq.~\eqref{eq:OU_app} which are obtained with integer combinations of the eigenvalues of $\mathbf{A}+\mathbf{B}$ \cite{metafunes2002}:
\begin{equation}
\mathrm{Re}(\lambda_\ell) = - \sum_{j=1}^{N/2} (n_j^+ + n_j^-) A_{jj},
\qquad
\mathrm{Im}(\lambda_\ell) = \sum_{j=1}^{N/2} (n_j^+ - n_j^-) \omega_j,
\end{equation}

Then, the double integral is
\begin{equation}
\int_0^t \int_0^t \chi_0(s-u) \mathrm{d}s \mathrm{d}u = 2\int_0^t(t-\tau)\chi_0(\tau)\mathrm{d}\tau
\approx 2 \sum_{\ell=1}^{M} \left[ \gamma_\ell I^{(c)}_\ell(t) + \xi_\ell I^{(s)}_\ell(t) \right],
\end{equation}
with
\begin{subequations}
\begin{align}
I^{(c)}_\ell(t) &= \int_0^t (t-\tau) e^{\mathrm{Re}(\lambda_\ell) \tau} \cos(\delta \mathrm{Im}(\lambda_\ell) \tau) \mathrm{d}\tau \\
&= \frac{ \mathrm{Re}(\lambda_\ell) (1 - e^{\mathrm{Re}(\lambda_\ell) t} \cos(\delta \mathrm{Im}(\lambda_\ell) t)) + \delta \mathrm{Im}(\lambda_\ell) e^{\mathrm{Re}(\lambda_\ell) t} \sin(\delta \mathrm{Im}(\lambda_\ell) t)}{ (\mathrm{Re}(\lambda_\ell))^2 + (\delta \mathrm{Im}(\lambda_\ell))^2 },\\
I^{(s)}_\ell(t) &= \int_0^t (t-\tau) e^{\mathrm{Re}(\lambda_\ell) \tau} \sin(\delta \mathrm{Im}(\lambda_\ell) \tau) \, \mathrm{d}\tau
\\
&= \frac{ \delta \mathrm{Im}(\lambda_\ell) (1 - e^{\mathrm{Re}(\lambda_\ell) t} \cos(\delta \mathrm{Im}(\lambda_\ell) t)) - \mathrm{Re}(\lambda_\ell) e^{\mathrm{Re}(\lambda_\ell) t} \sin(\delta \mathrm{Im}(\lambda_\ell) t)}{ (\mathrm{Re}(\lambda_\ell))^2 + (\delta \mathrm{Im}(\lambda_\ell))^2 }.
\end{align}
\end{subequations}
For large $\delta$, this implies that:
\begin{equation}\label{eq:order_denominator}
    2 \sum_{\ell=1}^{M} \left[ \gamma_\ell I^{(c)}_\ell(t) + \xi_\ell I^{(s)}_\ell(t) \right] \sim \mathcal{O}(\delta^{-2}),
\end{equation}
similar to what is obtained in Eq.~\eqref{eq:snr_D_delta}. The square root of the latter quantity gives the denominator of Eq.~\eqref{eq:ttcf_limit_epsilon} and scales with the inverse of $\delta^2$ for large values of $\delta$. Hence, accounting for Eq.~\eqref{eq:order_denominator}, we have:
\begin{equation}
    \mathrm{SNR}(C(t)) = \frac{\sum_{j=1}^N\frac{\kappa_j(t)}{\delta^2 \left(\left(\frac{\lambda_j}{\delta}\right)^2 + \omega_j^2\right)} \mathrm{d}s}{\sqrt{2 \sum_{\ell=1}^{M} \left[ \gamma_\ell I^{(c)}_\ell(t) + \xi_\ell I^{(s)}_\ell(t) \right]}} \sim \mathcal{O}(\delta^{-1}).
\end{equation}
This shows that the SNR ratio of the TTCF method is better than taking direct averages in the case of strong rotational forcing.

\section{Derivation of the TTCF relation for Markov chains}\label{TTCFMarkov}
 Let $\mathcal{X} = \{1,\dots,S\}$ be a finite state space and $\mathbf{M}$ an $S\times S$ Markov matrix acting on such space. We assume that the stochastic matrix $\mathbf{M}$ has a uniform contraction, i.e. for all probability vectors $\mathbf{u}$ and $\mathbf{v}$, the following applies:
\begin{equation}
\|\mathbf{M}\mathbf{u} - \mathbf{M}\mathbf{v}\|_{\mathrm{TV}} \le r \|\mathbf{u}-\mathbf{u}\|_{\mathrm{TV}}.
\end{equation}
where $0<r<1$ is the Dobrushin coefficient and $\|\mathbf{u}\|_{\mathrm{TV}} = \frac12 \sum_i u_i$. As a result, there is a unique ergodic invariant probability measure $\mathbf{u}_{inv}$ such that $\forall$ probability vectors $\mathbf{u}$ we have $\mathbf{u}_{inv} = \lim_{n\rightarrow\infty} \mathcal{M}^n\mathbf{u}$. Furthermore, $\mathcal{M}\mathbf{u}_{inv} =\mathbf{u}_{inv}$. 
The convergence to the limit sequence is exponential: $\| \mathcal{M}^n \mathbf{u} - \mathbf{u}_{inv}\|_{\mathrm{TV}} \le Cr^n$ \cite{Dobrushin1956central1,Dobrushin1956central2}. 

We now consider the perturbed stochastic matrix $\mathbf{M}_\varepsilon=\mathbf{M}+\varepsilon \mathbf{P}$, where $\mathbf{P}$ is a signed matrix such that $\mathbf{M}_\varepsilon$ is stochastic for sufficiently small $\varepsilon$. We assume that uniform contraction applies also for $\mathbf{M}_\varepsilon$, which then possesses a unique invariant measure $\mathcal{M}_\varepsilon\mathbf{u}_{inv,\varepsilon} =\mathbf{u}_{inv,\varepsilon}$. An observable vector $\boldsymbol{\Psi}$ in $\mathbb{R}^{S}$ has an expectation value with respect to the invariant vector $\mathbf{u}_{inv}$ given by:
\begin{equation}
    \langle \boldsymbol{\Psi} \rangle_{0} = \langle \boldsymbol{\Psi} , \mathbf{u}_{inv} \rangle = \sum_{i=1}^S\Psi_iu_{inv,i},
\end{equation}
Similarly, $\langle \boldsymbol{\Psi} \rangle_{\varepsilon} = \langle \boldsymbol{\Psi} , \mathbf{u}_{inv,\varepsilon} \rangle $.

If the perturbation matrix $\mathbf{P}$ is applied at time $n=0$, to the process distributed according to $\mathbf{u}_{inv}$, the expectation value of $\boldsymbol{\Psi}$ will be perturbed accordingly. Precisely, the expectation value $\langle \boldsymbol{\Psi} \rangle_{\varepsilon}(n)$, of $\boldsymbol{\Psi}$ at time $n$ is given by
\begin{equation}
\langle \boldsymbol{\Psi} \rangle_\varepsilon (n) = \langle \boldsymbol{\Psi},\mathbf{M}_\varepsilon^n\mathbf{u}_{inv}\rangle=\langle\left(\mathbf{M}_\varepsilon^\ast\right)^n\boldsymbol{\Psi},\mathbf{u}_{inv}\rangle.
\end{equation}
This gives us the direct formula for evaluating the response of the system: 
\begin{equation}
\langle \boldsymbol{\Psi} \rangle_\varepsilon (n)-\langle\Psi\rangle_0=\langle\left(\mathbf{M}_\varepsilon^\ast\right)^n\boldsymbol{\Psi},\mathbf{u}_{inv}\rangle-\langle\Psi\rangle_0,\label{directformulamarkov}
\end{equation}
which is, for Markov chains, the vectorized formulation of Eq.~\eqref{eq: response}. Moreover,
\begin{equation}
\langle \boldsymbol{\Psi} \rangle_\varepsilon (k+1)-\langle \boldsymbol{\Psi} \rangle_\varepsilon (k)=\langle\left(\mathbf{M}_\varepsilon^\ast\right)^k\boldsymbol{\Psi} , (\mathbf{M}_\varepsilon-\mathbf{M})(\mathbf{u}_{inv})\rangle= \varepsilon \langle\left(\mathbf{M}_\varepsilon^\ast\right)^k\boldsymbol{\Psi} : \boldsymbol{\Omega},\mathbf{u}_{inv}\rangle,
\end{equation}
where $\boldsymbol{\Omega}$ is the vectorized dissipation function defined, component-wise as:
\begin{equation}
    \Omega_i = \frac{1}{u_{inv,i}}\sum_{j=1}^S  P_{ij} u_{inv,j} ,
\end{equation}
for $i=1,\ldots,S$. Summing the expression over $k$ from 0 to $n$, we have: 
\begin{align}
\langle \boldsymbol{\Psi} \rangle_\varepsilon (n)-\langle \Psi\rangle_0&=\varepsilon\sum_{k=0}^n\langle\left(\mathbf{M}_\varepsilon^\ast\right)^k\Psi : \boldsymbol{\Omega},\mathbf{u}_{inv}\rangle\label{indirectformulamarkov},
\end{align}
which is the TTCF for Markov chains and corresponds  to Eq.~\eqref{eq:ttcf}.

\bibliographystyle{plain}
\bibliography{tesis}

\end{document}